\documentclass[aps,prl,twocolumn,superscriptaddress,showpacs]{revtex4}
\usepackage{graphicx}
\usepackage{dcolumn}
\usepackage{bm}

\newcommand{\rgn}{($\gamma$,n)}
\newcommand{\rng}{(n,$\gamma$)}

\newcommand{\taN}{$^{180}$Ta}
\newcommand{\taiso}{$^{180{\rm{m}}}$Ta}
\newcommand{\tags}{$^{180}$Ta$_{\rm{g.s.}}$}

\newcommand{\spro}{s-process}
\newcommand{\rpro}{r-process}
\newcommand{\ppro}{p-process}
\newcommand{\theff}{$t_{1/2}^{\rm{eff}}$}

\begin{document}

\title{
Survival of Nature's Rarest Isotope $^{180}$Ta under Stellar Conditions
}

\author{P.\ Mohr}
\email[E-mail: ]{WidmaierMohr@compuserve.de}
\affiliation{
Diakoniekrankenhaus Schw\"abisch Hall, D-74523 Schw\"abisch Hall,
Germany
}

\author{F.\ K\"appeler} 
\affiliation{
Forschungszentrum Karlsruhe, Institut f\"ur Kernphysik, P.O. Box 3640,
D-76021 Karlsruhe, Germany
}

\author{R.\ Gallino}
\affiliation{
Dipartimento di Fisica Generale, Universit{\`a} di Torino,
Via P.~Giuria 1, I-10125 Torino, Italy
}

\date{\today}

\begin{abstract}
The nucleosynthesis of nature's rarest isotope \taN\ depends
sensitively on the temperature of the astrophysical environment
because of depopulation of the long-living isomeric state via
intermediate states to the short-living ground state by thermal
photons. Reaction rates for this transition have been measured in the
laboratory. These ground state rates understimate the stellar rates
dramatically because under stellar conditions intermediate states are
mainly populated by excitations from thermally excited states in
\taiso . Full thermalization of \taN\ is already achieved for typical
\spro\ temperatures around $kT = 25$\,keV.  Consequently, for the
survival of \taN\ in the \spro\ fast convective mixing is required
which has to transport freshly synthesized \taN\ to cooler regions. In
supernova explosions \taN\ is synthesized by photon- or
neutrino-induced reactions at temperatures above $T_9 = 1$ in thermal
equilibrium; independent of the production mechanism, freeze-out from
thermal equilibrium occurs at $kT \approx 40$\,keV, and only $35 \pm
4$\,\% of the synthesized \taN\ survive in the isomeric state.
\end{abstract}


\maketitle

The nucleosynthesis of nature's rarest isotope and only naturally
occuring isomer \taN\ has been studied intensively over the last
years. In a simplified view, there seems to be no production path to
\taN\ in the slow and rapid neutron capture processes (\spro , \rpro
), and it has been pointed out that \taN\ is a natural product of the
\ppro\ without neutrino interactions \cite{Arn03,Uts03,Rau02} or with
neutrino interactions \cite{Heg05,Woo90}. $^{181}$Ta\rgn \taN\
photoproduction cross sections have been measured in
\cite{Uts03,Gok06}. But the \spro\ may also contribute to the
production of \taN\ via $\beta$-decay of thermally excited $^{179}$Hf
to $^{179}$Ta and subsequent $^{179}$Ta(n,$\gamma$)\taN\ neutron
capture \cite{Sch99}, 
or via $^{179}$Hf\rng $^{180{\rm{m}}}$Hf($\beta^-$)\taN\ \cite{Bee81}. 
Even the \rpro\ might produce \taN\ via the
$^{180{\rm{m}}}$Hf($\beta^-$)\taN\ decay \cite{Bee81,Yok83}; however,
the isomer decay branches in $A = 180$ nuclei are too small to explain
the stellar abundance of \taN\ \cite{Kel92,Les86,Kel86,Kel85}. The
astrophysical sites for the nucleosynthesis of \taN\ are still
uncertain. A combination of the above processes seems to be most
likely.

The \spro\ production
of \taN\ depends on the neutron capture cross sections of $^{179}$Ta
\cite{Sch99} 
and \taN\ \cite{Wis01,Wis04,Kae04} and the photodestruction of
the $9^-$ isomeric state in \taN\ which has been studied using real
photons from bremsstrahlung and from radioactive sources and virtual
photons in Coulomb excitation experiments
\cite{Bel99,Bel02,Lak00,Car89,Col88,Col90,Nem92,Nor84,Bik99,Sch94,Sch98,Loe96,Sch01,Loe03}.
It is the scope of the present paper to improve the analysis of the
above experimental data and to calculate the effective half-life of \taN\
under \spro\ conditions and the survival probability of \taN\ in
supernova explosions after production by photon- and neutrino-induced
reactions. The results are essential to explain the extremely low
abundances of \taN\ (isotopic: $0.01201 \pm 0.00008\,\%$; solar:
$2.49 \times 10^{-6}$ relative to Si $= 10^6$)
\cite{Lae05}. 

The effective half-life \theff\ of \taN\ depends sensitively on
temperature because the $1^+$ ground state of \taN\ has a short
half-life of only $t_{1/2,{\rm{g.s.}}} = 8.154$\,h, whereas the $9^-$
isomeric state ($J = 9$ confirmed very recently in \cite{Bis06})
at $E_x = 77.1$\,keV has a very long half-life
$t_{1/2,{\rm{iso}}} > 1.2 \times 10^{15}$\,yr; very recently, this
value has been improved by a factor of six to $t_{1/2,{\rm{iso}}} >
7.1 \times 10^{15}$\,yr \cite{Hul06}.
Photoexcitation from the
$9^-$ isomer to intermediate states (IMS) in the thermal photon bath
at energies around 25\,keV couples the long-living isomer to the
short-living ground state.

Realistic stellar models of \spro\ nucleosynthesis have shown that the
$^{13}$C($\alpha$,n)$^{16}$O and $^{22}$Ne($\alpha$,n)$^{25}$Mg
reactions are the neutron sources for the \spro . The
$^{13}$C($\alpha$,n)$^{16}$O reaction operates at thermal energies
around $kT \approx 8$\,keV for about $10^4 - 10^5$ years, but does not
contribute significantly to the \taN\ production. The higher energy of
$kT \approx 26$\,keV of the $^{22}$Ne($\alpha$,n)$^{25}$Mg reaction is
required for the production path via $\beta$-decay of thermally
excited $^{179}$Hf to $^{179}$Ta and subsequent
$^{179}$Ta(n,$\gamma$)\taN\ neutron capture. The
$^{22}$Ne($\alpha$,n)$^{25}$Mg neutron source operates during
convective helium shell flashes which last of the order of a few years
\cite{Gal98,Lug03,Str06}. Convective mixing transports
freshly synthesized \taN\ to cooler regions with a turnover time of
less than one hour \cite{Rei04}. Thus, for a
survival of \taN\ in the \spro\ the effective half-life of \taN\ has
to be ($i$) of the order of the mixing time scale at $kT = 26$\,keV to
survive the mixing in the flash and ($ii$) of the order of $10^5$
years at $kT = 8$\,keV to survive the \spro\ conditions between the
shell flashes. It will be shown that the second condition ($ii$) is
always fulfilled; however, the result of the present work shows that
convective mixing has to occur on timescales shorter than a few hours
-- as calculated in \cite{Rei04} --
to fulfill condition ($i$).

\begin{figure}[hbt]
\includegraphics[ bb = 105 90 465 343, width = 75 mm, clip]{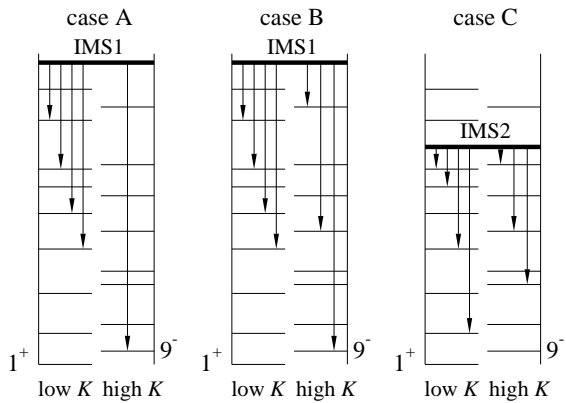}
\caption{
\label{fig:scheme} 
Simplified level scheme of \taN\ with low-$K$ (left) and high-$K$
(right) states and two IMS. The high-$K$ branch of the IMS1 may decay
directly to the $9^-$ isomer (case A, experimentally confirmed
\cite{Bel99,Bel02}) or via different branches (case B, not measured
yet). For another IMS2 the high-$K$ branch may decay only to excited
states of the high-$K$ bands (case C, not measured yet). The latter
case C will dramatically reduce the effective half-life of \taN\ under
\spro\ conditions
whereas case B leads only to a minor reduction of \theff\ but may
affect the freeze-out in the \ppro\ (see text).
}
\end{figure}

In a big number of gamma-spectroscopic studies no IMS in \taN\ has
been identified which mixes the $K$ quantum number and has
simultaneous $\gamma$-ray branches to the low-$K$ ground state and to
the high-$K$ isomeric state (see e.g.\
\cite{Dra98,Sai99,Dra00,Wen02}). However, at higher excitation
energies the $K$ quantum number is eroded.  It is clear that IMS must
exist because of the measured yield in the photoactivation
experiments. This is schematically shown in Fig.~\ref{fig:scheme}
where states with low $K$ (high $K$) decaying to the ground state
(isomeric) band are shown on the left (right) hand side.

\begin{figure}[hbt]
\includegraphics[ bb = 75 88 450 428, width = 75 mm, clip]{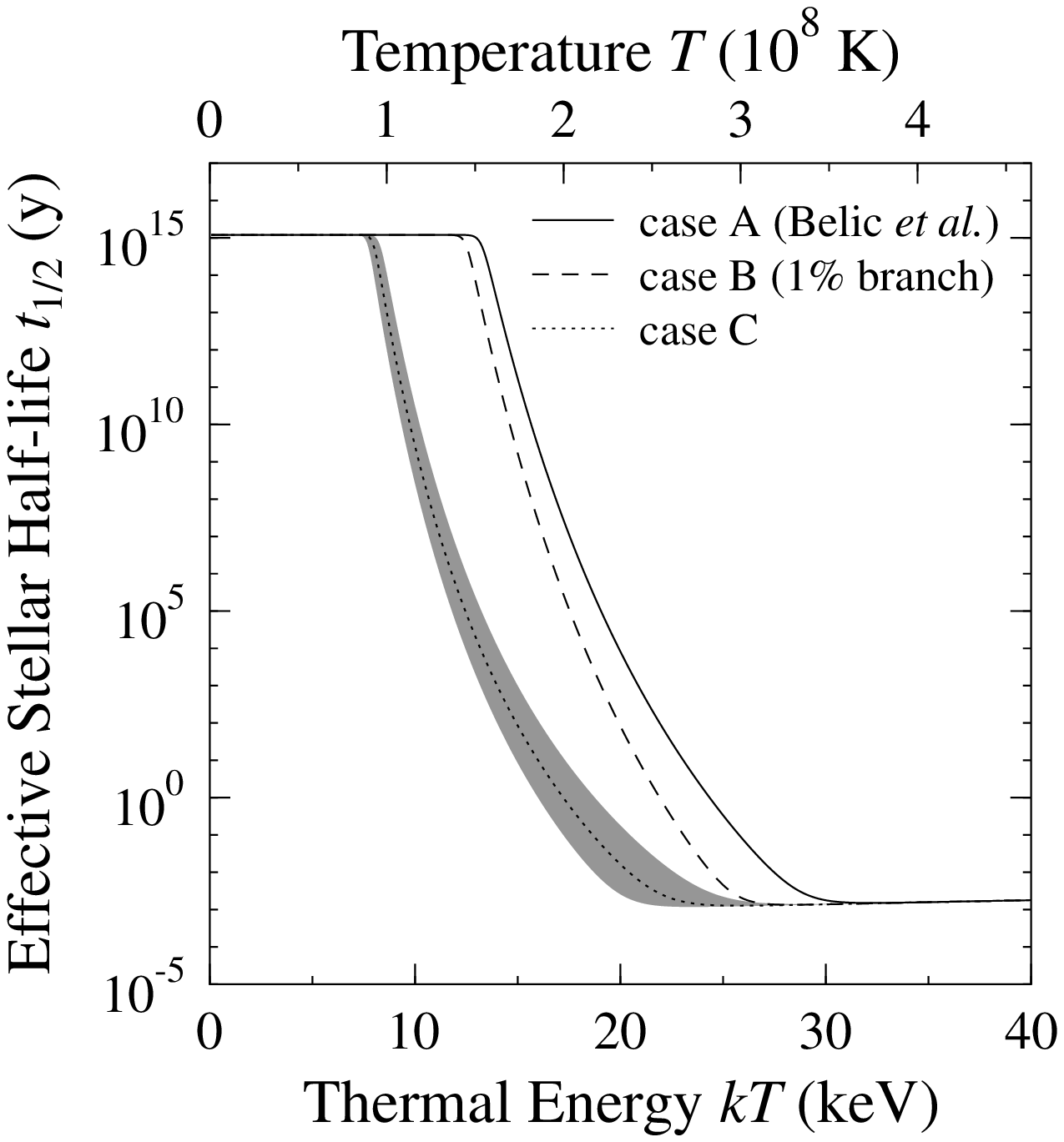}
\caption{
\label{fig:lambda} 
Effective stellar half-life $t_{1/2}$ of \taN\ calculated from the
half-lives of the $1^+$ ground state and $9^-$ isomer and the
integrated cross section $I_\sigma$ . The effective half-life of \taN\
at $kT = 26$\,keV is around 11\,hours (case C). For further discussion see
text.
}
\end{figure}

Photodestruction of \taiso\ has been identified down to an energy of
about 1010\,keV above the $9^-$ isomer \cite{Bel99,Bel02}, and an
energy-integrated cross section of $I_\sigma^{\rm{lab}} =
5.7$\,eV\,fm$^2$ for the transition from the isomer to the ground
state band has been derived. The experimental data do not exclude that
the IMS is located at lower energies and has a smaller cross
section. A tentative assignment for the IMS is given in \cite{Wal01}
with $J^\pi{\rm{(IMS)}} = (8^+)$ and $E_x = 1076$\,keV; this IMS
belongs to the $K^\pi = 5^+$ band with the band head at $E_x =
594$\,keV. The given $I_\sigma^{\rm{lab}}$ leads to the effective
half-life as shown in Fig.~\ref{fig:lambda}, case A.

Up to now, the effective half-life \theff\ has been calculated from a
three-level model taking into account the long half-life of the
isomeric state, the short half-life of the ground state, and the
integrated cross section $I_\sigma^{\rm{lab}}$ for the direct
transition from the $9^-$ isomeric state via the IMS to the ground
state band \cite{Bel99,Bel02}. However, the integrated cross section
has been measured under laboratory conditions. Under stellar
conditions the direct transition from the $9^-$ isomer to the IMS is
only a minor contribution to the stellar integrated cross section
$I^\ast_\sigma$; this has not been taken into account up to now.

It is a general feature of photon-induced reactions that cross
sections under stellar and laboratory conditions show noticeable
differences. The cross section in the laboratory for a resonance with
$J_R$ at $E_x$ in the reaction ($\gamma$,$X$) with the target in a
defined state with spin $J_0$ and excitation energy $E_0$ is given by
\begin{equation}
\sigma^{\rm{lab}}(E_\gamma) = \frac{g}{2} \, \frac{\pi}{k_\gamma^2} \,
\frac{\Gamma_\gamma^{J_R \rightarrow J_0} \, 
\Gamma_X}{[E_\gamma-(E_x-E_0)]^2 + \Gamma^2/4}
\label{eq:sigma}
\end{equation}
with the statistical factor $g = {(2J_R+1)}/{(2J_0+1)}$, the 
photon energy $E_\gamma = \hbar c k_\gamma$, the partial radiation
width $\Gamma_\gamma^{J_R \rightarrow J_0}$ for the direct transition from
$J_R$ to $J_0$, the partial width $\Gamma_X$ for the decay into the
$X$ channel, and the total width $\Gamma$. The energy-integrated cross
section is obtained by integration over $E_\gamma$:
\begin{equation}
I_\sigma^{\rm{lab}} = g \, \left(\frac{\pi \hbar c}{E_x - E_0}\right)^2
\, \frac{\Gamma_\gamma^{J_R \rightarrow J_0} \, \Gamma_X}{\Gamma}
\label{eq:isigmalab}
\end{equation}
The reaction rate $\lambda^{\rm{lab}}$ is given by
\begin{eqnarray}
\lambda^{\rm{lab}}(T) & = & c \int n_\gamma(E_\gamma,T) \,
\sigma^{\rm{lab}}(E_\gamma) \, dE_\gamma \nonumber \\
& \approx & c \, n_\gamma(E_x-E_0,T) \, I_\sigma^{\rm{lab}}
\label{eq:rate-lab}
\end{eqnarray}
with the thermal photon density
\begin{equation}
n_\gamma \, (E_\gamma,T) \, dE_\gamma = 
	\frac{1}{\pi^2} \, \frac{1}{(\hbar c)^3} \, 
	\frac{E_\gamma^2}{\exp{(E_\gamma/kT)} - 1} \, dE_\gamma
\label{eq:planck}
\end{equation}

Let us now for simplicity assume in the next paragraph that there is
one excited state with $J_1$ at energy $E_1$. The thermal population
ratio is given by
\begin{equation}
\frac{n_1}{n_0} = \frac{(2J_1+1)}{(2J_0+1)}
\exp{\left(-\frac{E_1-E_0}{kT}\right)}
\label{eq:boltz}
\end{equation}
and decreases exponentially with the factor $\exp{(-E_1/kT)}$. However, the
reaction rate $\lambda_1$ from this state is comparable to
$\lambda^{\rm{lab}}$ because the required photon energy $E_\gamma$ for
excitation of the resonance at $E_x$ is reduced by $E_1$ and thus the
photon density in Eq.~(\ref{eq:planck}) is enhanced by the same
exponential factor.

After summing over all properly weighted contributions from all thermally
excited states the integrated cross section $I_\sigma^\ast$ for one
IMS under stellar conditions is given by
\begin{equation}
I_\sigma^{\ast} = g \, \left(\frac{\pi \hbar c}{E_x - E_0}\right)^2 \,
\frac{\Gamma_\gamma^{\Sigma J_R \rightarrow J_i \rightarrow J_0} \,
\Gamma_X}{\Gamma}
\label{eq:isigmastar}
\end{equation}
with the partial radiation width $\Gamma_\gamma^{\Sigma J_R
\rightarrow J_i \rightarrow J_0}$ summed over all partial widths
$\Gamma_\gamma^{J_R \rightarrow J_i}$ leading finally to the initial
state with $J_0$. Note that the properties of the thermally excited
states with $J_i$ cancel out in Eq.~(\ref{eq:isigmastar}). Thermal
equilibrium is maintained by strong, i.e.\ mainly dipole and quadrupole,
transitions with $\Delta J = 1$ or 2; thermalization between states
with large $\Delta J$ is achieved by multiple transitions with low
$\Delta J$.

The ratio
between the reaction rates and integrated cross sections under stellar
and laboratory conditions is given by the branching ratio:
\begin{equation}
\frac{\lambda^\ast}{\lambda^{\rm{lab}}} =
  \frac{I_\sigma^\ast}{I_\sigma^{\rm{lab}}} =
  \frac{\Gamma_\gamma^{\Sigma J_R \rightarrow J_i \rightarrow J_0}}
  {\Gamma_\gamma^{J_R \rightarrow J_0} }
\label{eq:ratio}
\end{equation}
For the example of an isolated resonance in the
$^{20}$Ne($\gamma$,$\alpha$)$^{16}$O reaction the contributions of
thermally excited states have been calculated explicitly in
\cite{Mohr06}.

For the case of \taN\ this means that the integrated cross section is
dramatically enhanced under stellar conditions compared to the
laboratory results \cite{Bel99,Bel02}. The stellar integrated
cross section $I_\sigma^\ast$ is given by
\begin{equation}
I_\sigma^{\ast} = g \, \left(\frac{\pi \hbar c}{E_{\rm{IMS}} -
E(9^-)}\right)^2 \, \frac{\Gamma_\gamma^{\Sigma \, {\rm{IMS}}
\rightarrow 9^-} \, \Gamma_\gamma^{\Sigma \, {\rm{IMS}}
\rightarrow 1^+}}{\Gamma}
\label{eq:isigmastar-ta}
\end{equation}
with the summed partial widths $\Gamma_\gamma^{\Sigma \, {\rm{IMS}}
\rightarrow 9^-}$ to the $9^-$ isomer at $E(9^-) = 77.1$\,keV,
$\Gamma_\gamma^{\Sigma \, {\rm{IMS}} \rightarrow 1^+}$ to the $1^+$
ground state, and the total width $\Gamma = \Gamma_\gamma^{\Sigma \,
{\rm{IMS}} \rightarrow 9^-} + \Gamma_\gamma^{\Sigma \, {\rm{IMS}}
\rightarrow 1^+}$. The effective stellar half-life under \spro\
conditions is mainly defined by the lowest IMS.

Let us first show the influence of various decay branches of the
experimentally known IMS1 
via excited states to the $9^-$ isomer. Case A in
Figs.~\ref{fig:scheme} and \ref{fig:lambda} repeats the result of
\cite{Bel99,Bel02} and takes into account only the direct decay from
IMS1 to the $9^-$ isomer.
If one makes the
hypothetical assumption that the direct decay of the ISM1 to the $9^-$ isomer
contributes only with 1\,\% to the total decay to the isomer
(case B in Figs.~\ref{fig:scheme} and \ref{fig:lambda}), one can
calculate $I_\sigma^\ast = 100 \, I_\sigma^{\rm{lab}}$ and a reduced
effective stellar half-life as shown in Fig.~\ref{fig:lambda} (dashed
line). Note that almost all states in \taN\ do not directly decay to
the ground state or to the $9^-$ isomer but via $\gamma$-ray cascades
(see adopted level scheme and adopted $\gamma$-rays in \cite{ENSDF}).

In the following paragraph we try to identify the lowest IMS which
defines the effective half-life of \taN\ under \spro\ conditions. The
experimental results for IMS in \cite{Bel99,Bel02} have been 
tentatively assigned to members of an intermediate $K$ band with
$K^\pi = 5^+$ and the band head at $E_x = 594$\,keV \cite{Wal01}. A
careful inspection of the level scheme of \taN\ in \cite{ENSDF} shows
that this band head is an excellent candidate for the lowest IMS (IMS2
in Fig.~\ref{fig:scheme}). The
half-life of this $5^+$ isomer is $t_{1/2} = 16.1 \pm 1.9$\,ns
\cite{ENSDF}, and it decays via 72\,keV M1 (or E2) decay to the $4^+$
state at $E_x = 520$\,keV. The half-life of the $5^+$ isomer
corresponds to $3.7 \times 10^{-3}$\,W.u. (M1) or about 300\,W.u.\
(E2). Let us now assume that there is a weak 1\,\% branching of this
$5^+$ isomer to the $7^+$ state at $E_x = 357$\,keV which is the band
head of the $K^\pi = 7^+$ band and decays via two subsequent
$\gamma$-transitions to the $9^-$ isomer at $E_x = 77.1$\,keV. 
Such E2 transitions with $\Delta K = 2$ have indeed been observed in
\taN\ and also in neighboring $^{176}$Lu \cite{ENSDF}.
The direct
transition from the $5^+$ state to the $9^-$ isomer (M4 or E5) is
extremely weak. This is illustrated as case C in
Fig.~\ref{fig:scheme}.

The E2 transition from the $5^+$ to the $7^+$ state with a 1\,\%
branching corresponds to $7.5 \times 10^{-3}$\,W.u.\ which is of the
same order of magnitude as the M1 transition. Such a transition leads
to an integrated stellar cross section of $I_\sigma^\ast = 2.3 \times
10^{-4}$\,eV\,fm$^2$. The effective stellar half-life is dramatically
reduced (see case C, shown as dotted line in
Fig.~\ref{fig:lambda}). The grey shaded error bar in
Fig.~\ref{fig:lambda} is obtained if one increases or reduces the E2
transition strength by a factor of 10. The assumed lower limit of less
than $10^{-3}$\,W.u.\ for a E2 transition from a $K^\pi = 5^+$ to a
$K^\pi = 7^+$ band is much lower than typical hindrance factors for
such a $\Delta K = 2$ transition. Even larger transition strengths
have been reported for a direct E2 or E3 transition from the $9^-$
isomer to an unidentified IMS in a Coulomb excitation experiment
\cite{Sch94}.

The 72\,keV decay of the $5^+$ level at $E_x = 594$\,keV has not been
confirmed in \cite{Wen02}. It is interesting to note that the results
for the effective stellar half-life of \taN\ are not affected if the
$5^+$ state decays preferentially by the above discussed E2 transition
to the $7^+$ state -- provided that there is a 1\,\% branching to
states which decay finally to the ground state because of the
symmetric roles of the $9^-$ isomer and $1^+$ ground state in the
stellar integrated cross section $I_\sigma^\ast$ in
Eq.~(\ref{eq:isigmastar-ta}). 

At the relevant thermal energy of $kT = 26$\,keV one finds that the
effective stellar half-life of \taN\ is about 11\,h from the above
mentioned model which includes the $1^+$ ground state, the $9^-$
isomer, and the IMS which connects the ground state and the
isomer. This short half-life is no longer defined by the transition
rate from the isomer via the IMS to the ground state (as assumed in
all previous work). Instead, such short half-lives correspond to full
thermalization of \taN\ at the relevant temperature, and the effective
half-life can simply be calculated from the properly weighted decay
rates of low-lying states in \taN . At $kT = 26$\,keV one finds about
57\,\% of \taN\ in the $1^+$ ground state, 21\,\% in the $2^+$ state
at $E_x = 40$\,keV, 19\,\% in the $9^-$ isomer at $E_x = 77$\,keV, and
about 3\,\% in higher-lying states. The half-lives of the $1^+$ ground
state and the $9^-$ isomer are well-known experimentally. However, the
$\beta$-decay half-life of the $2^+$ state is experimentally unknown
because the $2^+$ state decays by M1 or E2 $\gamma$-rays to the $1^+$
ground state. It has been estimated in \cite{Tak87} that thermal
population of low-$K$ states slightly increases the half-life of \taN
. A further increase of the effective half-life of \taN\ by up to a
factor of three at temperatures around $kT = 10 - 30$\,keV comes from
the ionization at high temperatures which reduces the electron capture
probability of \tags\ \cite{Tak87,Nor84}.

It has been shown recently \cite{Rei04} that the convective turnover
time in helium shell flashes of AGB stars is of the order of one hour,
and the time required for the transport of freshly synthesized nuclei
to cooler regions may even be shorter of the order of
minutes. Consequently, a reasonable amount of \taN\ can survive under
\spro\ conditions. A precise calculation of the \spro\ production of
\taN\ requires the knowledge of production and destruction cross
sections of the ground state and the $9^-$ isomer in \taN\ and coupling
of ground state and $9^-$ isomer via the IMS \cite{Kae04}. These
nuclear physics ingredients have to be combined with a time-dependent
model of the helium shell flashes in AGB stars. 

As an alternative to the \spro\ nucleosynthesis, photon-induced
reactions or neutrino-induced reactions during type II supernova
explosions have been suggested for the production of \taN\
\cite{Arn03,Uts03,Rau02,Heg05,Woo90,Gok06}. For photon-induced
reactions the lower end of the adopted temperature region around $T_9
= 2 - 3$ is required because of possible production and destruction by
$^{181}$Ta\rgn \taN \rgn $^{179}$Ta reactions. Neutrino nucleosynthesis of
\taN , mainly by the charged current reaction
$^{180}$Hf($\nu_e$,$e^-$)\taN , occurs also at high temperatures above
$T_9 = 1$ \cite{Heg05,Pru05,Pru05a}. The stellar reaction rates
$\lambda^\ast$ for the transition from the isomer to the ground state
and its inverse are linked together by detailed balance and exceed
$\lambda^\ast = 10^5$/s at $T_9 = 1$; thus, \taN\ is completely
thermalized independent of the production mechanism by photons or
neutrinos. As soon as the reaction rate $\lambda^\ast$ drops below the
time scale of the supernova explosion, \taN\ freezes out in the
thermal composition at freeze-out temperature $kT_{\rm{f}}$. We
calculate the freeze-out temperature $kT_{\rm{f}} = 40.4 \pm 5.5$\,keV
from the reaction rate $\lambda^\ast = 1$/s; the uncertainties are
carefully estimated using $\lambda^\ast = 10$/s (0.1/s) for the upper
(lower) limit of $kT_{\rm{f}}$. Taking into account all known levels
of \taN, one finds that $34.8^{+3.6}_{-4.6}$\,\% of the synthesized
\taN\ is in states with high $K$ at $kT_{\rm{f}}$ and survives finally
in the $9^-$ isomer. This is at the lower end of previous estimates of
about 30\,\%$-$ 50\,\% for the survival probability
\cite{Rau02,Arn03}.

It is interesting to note that the stellar reaction rate
$\lambda^\star$ above $T_9 > 1$ is dominated by the properties of
the lowest experimentally known IMS at $E_x = 1076$\,keV from
\cite{Bel99,Bel02} and further higher-lying IMS. Thus, full
thermalization of \taN\ under \ppro\ conditions is inevitable and
based on experimental data. The reaction rate $\lambda^\star(594)$ of
the new IMS at $E_x = 594$\,keV is the same as $\lambda^\star(1076)$ 
at a temperature of $T_9 = 0.500$ or $kT = 43.1$\,keV: $\lambda^\star(594) =
\lambda^\star(1076) \approx 1.5/s$. The new IMS at $E_x =
594$\,keV shifts the freeze-out temperature $kT_f$ only slightly down by
about 2\,keV and reduces the survival probability of the $9^-$ isomer
from 36.1\,\% to 34.8\,\%. A slightly stronger shift of $kT_f$ is
found for the hypothetical case B in Fig.~\ref{fig:scheme}.

In conclusion, it has been shown that thermal excitation from the
long-living $9^-$ isomer in \taN\ leads to an enhanced stellar cross
section $I_\sigma^\ast$ for transitions from the $9^-$ isomer to the
$1^+$ ground state via an intermediate state. At $kT = 26$\,keV
which is required for the \spro\ production of \taN\ we find that
\taN\ is fully thermalized. Fast convective mixing is required for the
survival of \taN\ because of the short effective half-life of
$t_{1/2}^{\rm{eff}} = 11$\,h. In supernovae \taN\ is produced above
$T_9 = 1$ in thermal equilibrium. Freeze-out at $kT \approx 40$\,keV
leads to survival of 35\,\% of the synthesized \taN\ independent of
the production mechanism by photons or neutrinos.

\begin{acknowledgments}
We thank T.\ Rauscher and K.\ Takahashi for encouraging discussions.
\end{acknowledgments}

\end{document}